\documentclass[journal=jctcce,manuscript=article,layout=twocolumn]{achemso}
\usepackage[version=3]{mhchem} 
\usepackage{physics}
\usepackage[T1]{fontenc} 
\usepackage[utf8x]{inputenc}

\usepackage{listings}
\usepackage{color, colortbl}
\usepackage[table]{xcolor}
\usepackage{float}

\definecolor{mygreen}{rgb}{0,0.6,0}
\definecolor{mygray}{rgb}{0.5,0.5,0.5}
\definecolor{mymauve}{rgb}{0.58,0,0.82}
\definecolor{python_bg}{RGB}{247, 247, 247}
\definecolor{halfgray}{gray}{0.55}
\definecolor{python_frame}{RGB}{207, 207, 207}

\author{Maximilian H. Kriebel}
\email{maximilian.kriebel@web.de}
\author{Pawe\l{} Tecmer}
\author{Marta Ga\l{}ynska}
\author{Aleksandra Leszczyk}
\author{Katharina Boguslawski}
\email{k.boguslawski@fizyka.umk.pl}
\affiliation[University]
{Institute of Physics, Faculty of Physics, Astronomy, and Informatics, Nicolaus Copernicus University in Toruń, Grudziadzka 5, 87-100 Toruń, Poland}

\title[Accelerating Pythonic coupled cluster implementations]
  {Accelerating Pythonic coupled cluster implementations: a comparison between CPUs and GPUs}

\abbreviations{pCCD, PyBEST, GPU, \texttt{CuPy}}
\keywords{pair coupled cluster doubles, geminals, Python, \texttt{CuPy}, Cholesky,  \LaTeX}
\makeatletter
\let\l@addto@macro\relax
\makeatother
\usepackage[fontsize=11pt]{scrextend}

\let\oldmaketitle\maketitle
\let\maketitle\relax
\begin{document}


\twocolumn[
\begin{@twocolumnfalse}
\oldmaketitle
\begin{abstract}
We scrutinize how to accelerate the bottleneck operations of Pythonic coupled cluster implementations performed on a \texttt{NVIDIA} Tesla V100S PCIe 32GB (rev 1a) Graphics Processing Unit (GPU).
The \texttt{NVIDIA} Compute Unified Device Architecture (CUDA) API is interacted with via \texttt{\texttt{CuPy}}, an open-source library for Python, designed as a \texttt{NumPy} drop-in replacement for GPUs.
The implementation uses the Cholesky linear algebra domain and is done in {PyBEST}, the Pythonic Black-box Electronic Structure Tool---a fully-fledged modern electronic structure software package. Due to the limitations of Video Memory (VRAM), the GPU calculations must be performed batch-wise.
Timing results of some contractions containing large tensors are presented. The \texttt{\texttt{CuPy}} implementation leads to factor 10 speed-up compared to calculations on 36 CPUs.
Furthermore, we benchmark several Pythonic routines for time and memory requirements to identify the optimal choice of the tensor contraction operations available.
Finally, we compare an example CCSD and pCCD-LCCSD calculation performed solely on CPUs to their CPU--GPU hybrid implementation.
Our results indicate a significant speed-up (up to a factor of 16 regarding the bottleneck operations) when offloading specific contractions to the GPU using \texttt{CuPy}.
\end{abstract}
\end{@twocolumnfalse}
]

\section{Introduction}
A critical job of a Graphics card is to compute projections of 3-dimensional objects to a 2D surface using linear algebra.
These calculations can be performed in parallel very effectively, meaning that multiple small mathematical operations like multiplication and addition can be performed simultaneously.
For this reason, Graphics Processing Unit (GPU) development mainly focuses on massively increasing the number of computing cores.
While a Central Processing Unit (CPU) may have up to 128 cores on the high end, GPUs already have up to~16,000 cores (like \texttt{NVIDIA} RTX 4090).

In scientific computing, the distribution of calculations among multiple CPU cores and multiple nodes is a standard practice.~\cite{gamess-prallelization-1998, arnim_98,mpqc-parallel-valeev-jpca-2016, qunatum-package2.0-jctc-2019,nwchemex-cr-2021}
Due to its inherently parallel structure, linear algebra operations can be calculated in parallel and, therefore, efficiently offloaded to the GPU.~\cite{martinez-gpu-review-2008, eri-gpu-matinez-jctc-2011, deprince-ccsd-gpu-jctc-2011, portion-fragmentation-gordon-jctc-2023, goetz-nvidia-amd-comparision-2023}
The first to report independently that the internal hugely parallel structure of GPUs can be misused, not to compute graphics, but to be utilized in quantum chemistry were Yasuda~\cite{gpu-eri-yasuda-jcc-2008} and Ufimtsev and Martinez,\cite{gpu-martinez-2008} respectively.
Today, the potential of GPUs for non-graphics-related computations is widely understood and often used to accelerate quantum chemistry methods like density functional approximations~\cite{yasuda-dft-gpu-2008}, Hartree--Fock theory~\cite{gpu-hf-gordon-jctc-2012, scf-gpu-gordon-jctc-2021}, Møller--Plesset perturbation theory \cite{gpu-mp2-guzik-jcpa-2008, gpu-mp2-martinez-2016, gpu-ri-mp2-gordon-jcp-2023}, coupled cluster theory~\cite{deprince-ccsd-gpu-jctc-2011, gpu-ccsd_t_jctc-2011, gpu-cc-gordon-jctc-2013, mpqc-parallel-valeev-jpca-2016, gpu-cc-dirac-jctc-2021}, and the evaluation of effective core potentials \cite{gpu-corepotential-martinez-2015} to name a few.
The \texttt{NVIDIA} CUDA API offers a \texttt{C++} interface to utilize GPU computing power.~\cite{gpu-high-throughput-bio, gpu-eri-gordon-mp-2023}
A relatively quick way to interact with this API is, for example, \texttt{CuPy}~\cite{cupy_paper}, a {Python} library which internally uses CUDA routines. 
\texttt{CuPy} brings a lot of attention from Python-based software developers as its interface is highly compatible with \texttt{NumPy}~\cite{numpy_paper-nature-2020} and allows even for a drop-in replacement in some particular cases.
Although sacrificing some fine control when exploiting libraries like \texttt{CuPy}, third-party libraries are quickly accessible without the necessity of in-detail back-end control and, therefore, a very convenient and efficient way to probe graphics processor utilization in the first place.

In this work, we will present benchmark results comparing the timings of the bottleneck tensor contractions present in coupled-cluster calculations restricted to at most double excitations, where Cholesky vectors approximate the electron repulsion integrals.~\cite{cholesky-review-2011} 
Specifically, we propose several flavors of performing these bottleneck contractions on a CPU using Pythonic libraries and benchmark their resource requirements.
These contractions are calculated with, for instance, \texttt{NumPy}'s~\cite{numpy_paper-nature-2020} \texttt{tensordot} and \texttt{einsum} routines or \texttt{opt\_einsum}~\cite{opt_einsum-2018} on CPUs.
Finally, we elaborate on exporting these operations on a GPU exploiting \texttt{CuPy}'s \texttt{tensordot} routine.~\cite{cupy_paper}

This work is organized as follows.
In section~\ref{sec:cc}, we briefly discuss the main bottleneck operations in coupled-cluster calculations.
Section~\ref{sec:python-cc-vector} scrutinizes several Python-based strategies to compute the coupled-cluster vector function. 
In section~\ref{sec:pybest-tce}, we examine the PyBEST tensor contraction engine.
A GPU implementation exploiting the \texttt{CuPy} library is summarized in section~\ref{sec:cupy}.
Numerical results and assessment of the GPU to CPU performance are presented in section~\ref{sec:cpu-gpu-comparison}. 
Finally, we conclude in section~\ref{sec:conclusions}.

\section{The CC ansatz}\label{sec:cc}
Our starting point is the coupled cluster (CC) ansatz~\cite{cizek_jcp_1966, cizek_paldus_1971, bartlett_rev_1981, bartlett_2007, helgaker_book, shavitt_book},
\begin{equation}
    \ket{\Psi} = e^{\hat T} \ket{\Phi_0}
\end{equation}
where $\hat T$ is the cluster operator and $\ket{\Phi_0}$ some reference wave function like the Hartree--Fock determinant.
In this work, we will consider, at most, double excitations in the cluster operator, that is, $\hat T = \hat T_2$.
We do not consider single excitations explicitly as the bottleneck operations are due to the $\hat T_2$ excitation operator.
Furthermore, we will work in the spin-free representation, with spin-free amplitudes, and the CC equations are spin summed.
In this picture, the double excitation operator takes on the form 
\begin{equation}
    \hat T_2 = \frac{1}{2} \sum_{ij}^{\rm occ}\sum_{ab}^{\rm virt} t_{ij}^{ab} \hat E_{ai} \hat E_{bj}
\end{equation}
with the CCD amplitude $ t_{ij}^{ab}$ and $\hat E_{ai}$ being the singlet excitation operator,
\begin{equation}
    \hat E_{ai} = \hat a^\dagger \hat i + \hat {\bar a}^\dagger \hat {\bar i},
\end{equation}
where $p^\dagger$ ($\bar p^\dagger$) indicates electron creation operators for $\alpha$ ($\beta$) electrons, while $p$ ($\bar p$) are the corresponding annihilation operators.
The above sum runs over all occupied (occ) and virtual (virt) orbitals for the chosen reference determinant $\ket{\Phi_0}$.

The scaling-determining step in the CCD amplitude equations is associated with the following term
\begin{equation}
    0 = \ldots + \sum_{cd} \bra{ab}\ket{cd} t_{ij}^{cd} + \ldots
\end{equation}
Summation over the indices $i,j,a,b$ is implied, which results in the formal scaling of the CCD equations of $\mathcal{O}(o^2v^4)$.
In the above equation, $\bra{ab}\ket{cd}$ are the electron-repulsion integrals in Physicist's notation.
To reduce the storage of the full electron-repulsion integrals, they can be approximated by Cholesky decomposition,~\cite{cholesky-koch-jcp-2003, cholesky-review-2011}
\begin{equation}\label{eq:cholesky}
    \bra{ab}\ket{cd} \approx \sum_{x} L_{ac}^x L_{bd}^x,
\end{equation}
where $x$ indicates the summation over the elements of the Cholesky vectors.
Since we work with real orbitals with 8-fold permutational symmetry of the electron-repulsion integrals (ERI), both Cholesky vectors $L_{ac}^x$ and $L_{bd}^x$ are identical.
Then, the CC excitation amplitudes are optimized iteratively by rewriting the rate-determining step of the vector function $\tilde{t}_{ij}^{ab}$ of iteration $n$ of the optimization procedure,
\begin{equation}\label{eq:vfunction}
    \tilde{t}_{ij}^{ab} = \ldots + \sum_{xcd} L_{ac}^x L_{bd}^x t_{ij}^{cd} + \ldots,
\end{equation}
including a third summation index running over the Cholesky vectors (ignoring the summation over the fixed indices $i,j,a,b$).
Thus, formally, the complexity increases to $\mathcal{O}(xo^2v^4)$.
Note that the dimension of $x$ depends on the chosen threshold of the Cholesky decomposition.
For decent to tight thresholds (around $10^{-6}$), we have $x \approx 5(o+v)$.

\section{Pythonic implementations of the CC vector 
function}\label{sec:python-cc-vector}
To solve for the CC amplitudes iteratively, we must evaluate the vector function of eq.~\eqref{eq:vfunction} in each iteration, including all its terms.
Thus, we will focus on the bottleneck operations of the corresponding CCD vector function.
Within a Pythonic implementation, we can utilize various Python libraries to perform summations efficiently without resorting to a nested for-loop implementation.~\cite{psi4numpy, pybest-paper, pyscf-jcp-2020, ghent-qc, gator-wires-2021, fanpy-jcc-2023}
Based on the chosen routines, these summations (in the following called contractions) can be performed in one shot or sequentially, creating several intermediates to boost efficiency and reduce resource requirements or even enabling out-of-the-box parallelization.
In the following, we will scrutinize different variants to evaluate eq.~\eqref{eq:vfunction} as pythonically as possible, focusing solely on Python features and libraries.
Specifically, we focus on the \texttt{NumPy} routines \texttt{einsum} and \texttt{tensordot}, the \texttt{opt\_einsum} package, and a GPU implementation exploiting \texttt{CuPy}'s \texttt{tensordot} feature.
\subsection{\texttt{einsum} and \texttt{opt\_einsum}}
The possibly easiest and most straightforward way of avoiding nested for-loop implementations when dealing with tensor contractions is to refer to \texttt{NumPy}'s \texttt{einsum} routine, which evaluates the Einstein summation convention on a sequence of operands.
Using \texttt{numpy.einsum}, many---albeit not all---linear algebraic operations on multi-dimensional arrays can be represented in a simple and intuitive language.
With increasing version number, additional features and improvements have been incorporated into the \texttt{numpy.einsum} function.
One crucial parameter is the \texttt{optimize} argument, which allows control over intermediate optimization.
If set to \texttt{"optimal"}, an optimal path of the contraction in question will be performed.
Another possibility is to exploit the \texttt{numpy.einsum\_path} function to steer the order of the individual contractions in the most optimal way.

An effort to improve the performance of the original \texttt{numpy.einsum} routine lead to the development of the \texttt{opt\_einsum} package.~\cite{opt_einsum-2018}
It offers several features to optimize \texttt{numpy.einsum} significantly, reducing the overall execution time of einsum-like expressions.
For instance, it automatically optimizes the order of the underlying contraction and exploits specialized routines or BLAS~\cite{blas-paper}.
\texttt{opt\_einsum} can also handle various arrays, like \texttt{NumPy}, \texttt{Dask}, \texttt{PyTorch}, \texttt{Tensorflow}, or \texttt{CuPy}, to name a few.
Furthermore, the optimization of \texttt{numpy.einsum} has been passed upstream to the original \texttt{numpy.einsum} project.
Some of \texttt{opt\_einsum}'s features can hence be utilized by \texttt{numpy.einsum} modifying the \texttt{optimization} option.
Since \texttt{opt\_einsum} features more up-to-date algorithms for complex contractions, we will focus on the \texttt{opt\_einsum.contract} function to evaluate the Einstein summation convention on a sequence of operands, typically containing three multi-dimensional input arrays.

\texttt{opt\_einsum.contract} represents a replacement for \texttt{numpy.einsum} where the order of the contraction is optimized to reduce the overall scaling (and hence increase the computational speed-up) at the cost of several intermediate arrays.
To steer the memory limit and prevent the generation of too large intermediates, \texttt{opt\_einsum.contract} offers the \texttt{memory\_limit} parameter to provide an upper bound of the largest intermediate array built during the tensor contraction.

Thus, the bottleneck contraction in eq.~\eqref{eq:vfunction} can be straightforwardly evaluated using, for instance, \texttt{opt\_einsum.contract} as follows
\begin{lstlisting}[language=Python]
    t_new[:] += opt_einsum.contract("xac,xbd,icjd->iajb", L_0, L_1, t_old)
\end{lstlisting}
In the above code snippet, \texttt{t\_new} indicates the vector function of iteration $n$, \texttt{t\_old} the current approximate solution of the CCD amplitudes, \texttt{L\_0} (\texttt{L\_1}) is the Cholesky vector of eq.~\eqref{eq:cholesky}.
Note that $\tilde{t}_{ij}^{ab}$ are stored as a 4-dimensional \texttt{NumPy} array \texttt{t\_new[i,a,j,b]}.

\subsection{\texttt{tensordot}}
An alternative routine to perform a tensor contraction is the \texttt{tensordot} function offered in \texttt{NumPy}\cite{numpy-developers-blog} and \texttt{CuPy}\cite{cupy_paper}.
It efficiently computes the summation of one (or more) given index (indices).
It allows for saving memory by de-allocating intermediate arrays and explicitly defining the path of the complete tensor contraction.
Furthermore, \texttt{tensordot} makes use of the BLAS\cite{blas-paper} API and features a multithreaded implementation when linked against the proper libraries like OpenBLAS,~\cite{openblas} MKL,~\cite{mkl} or ATLAS.~\cite{atlas-paper}
A contraction along one axis of two arrays \texttt{A} and \texttt{B},
\begin{equation*}
    C_{ij} = \sum_{n} A_{ni} B_{nj},
\end{equation*}
translates into
\begin{lstlisting}[language=Python]
C = tensordot(A, B, axes=([0, 0]))
\end{lstlisting}
Similarly, \texttt{tensordot} can contract (that is, sum over) two or more axes in one function call, where
\begin{equation*}
    C_{ij\ldots kl} = \sum_{nm\ldots} A_{nimj\ldots} B_{nklm\ldots},
\end{equation*}
translates into
\begin{lstlisting}[language=Python]
C = tensordot(A, B, axes=([0,1], [0,2]))
\end{lstlisting}
Note, however, that \texttt{tensordot} allows for contracting only two operands at a time.
Thus, to evaluate the term in eq.~\eqref{eq:vfunction}, a sequence of \texttt{tensordot} calls must be performed where suitable intermediates are created.
One possibility is to contract the Cholesky vectors to create an intermediate of dimension $v^4$, which is then passed to a second \texttt{tensordot} call generating the desired output,
\begin{lstlisting}[language=Python]
# first intermediate generates the dense ERI[a,c,b,d]
eri_acbd = tensordot(L_0, L_1, axes=([0, 0]))
# second contraction results in t_new[i,j,a,b]
t_new[:] += tensordot(t_old, eri_acbd, axes=([1,3], [1,3])).transpose((0,2,1,3))
\end{lstlisting}
Note that \texttt{tensordot} does not reorder the axis.
In that case, we need to transpose the intermediate result to match the shape of the output array (the vector function).
However, generating a $v^4$ intermediate of the ERI might be prohibitive regarding memory requirements for larger systems.
Other possibilities lead to even larger intermediates.
For instance, contracting \texttt{L\_0} with \texttt{t\_old} yields a multi-dimensional array of size $x o^2v^3$, which is smaller than the \texttt{eri\_acbd} intermediate if $xo^2 \ll v$, where $x$ is a pre-factor depending on the threshold of the Cholesky-decomposed ERI.
This pre-factor is typically challenging to determine \textit{a priori}.
Nonetheless, for computationally feasible problems, the condition $ao^2 \ll v$ is rarely satisfied, making the first contraction path computationally more efficient in terms of memory.

A less elegant, albeit computationally cheaper way to use all the benefits of the \texttt{tensordot} function is to introduce one for-loop to iterate over one axis.
If we choose to loop over the second axis of \texttt{L\_0}, we generate intermediates of at most $v^3$,
\begin{lstlisting}[language=Python]
for a in range(L_0.shape[1]):
    # first contract two Cholesky vectors
    eri_bdc = numpy.tensordot(L_1, L_0[:, a, :], axes=([0, 0]))
    # contract t_old[i,c,j,d] with eri_bdc (sum over c and d)
    t_new[:, a, :, :] += tensordot(t_old, eri_bdc, axes=([1, 3], [2, 1]))
\end{lstlisting}
The following will refer to the contraction path above as our \texttt{numpy.tensordot} routine.
Note, however, that this is not purely a \texttt{numpy.tensordot} computation, but an iterative call of the \texttt{numpy.tensordot} method to calculate eq.~\eqref{eq:vfunction} to prevent the creation of $v^4$ intermediate tensors.

\section{A modular implementation of tensor contractions}\label{sec:pybest-tce}
We have implemented and benchmarked the performance of the above-mentioned tensor contraction routines in PyBEST.~\cite{pybest-paper}
Specifically, PyBEST is designed as a modular toolbox where the wave-function-specific implementations are decoupled from the linear algebra operations.
In an actual calculation, the logic in choosing the optimal tensor-contraction scheme is as follows
\begin{lstlisting}[language=Python]
    try:
        cupy-accelerated contraction
    except NotImplementedError:
        try:
            numpy.tensordot contraction
        except ArgumentError:
            opt_einsum if not available then numpy.einsum(..., optimize="optimal")
\end{lstlisting}
The first \texttt{try} statement enforces that selected tensor contractions are performed on the GPU if available.
If bottleneck-specific contractions are not supported, or a CUDA-ready GPU is unavailable, a \texttt{numpy.tensordot} call is performed.
Since \texttt{numpy.tensordot} supports only specific contractions featuring non-repeating indices (that is, repeated indices have to be summed over in \texttt{numpy.tensordot}), an \texttt{opt\_einsum} call is performed or, if \texttt{opt\_einsum} is not available, an optimized \texttt{numpy.einsum} function call is made instead.
The corresponding tensor contraction operation is written using the Einstein-summation convention of the \texttt{numpy.einsum} module, that is, all mathematical operations use an input and output string, where repeated indices are summed over.
That allows for one unique notation of mathematical operations independent of the underlying representation of the tensors, especially the ERI.
As an example, the bottleneck contraction in eq.~\eqref{eq:vfunction} translates to the string \texttt{"abcd,ecfd->eafb"}, where the first part (\texttt{abcd}) corresponds to the ERI, the second part (\texttt{ecfd}) to the doubles amplitudes, while the output string (\texttt{eafb}) indicates the order of the output indices of the vector function.
Internally, this string is further decoded according to the notation used in the employed \texttt{LinalgFactory} instance, a dense or Cholesky-decomposed representation.
If a dense representation of tensors is chosen, the string remains as is.
For Cholesky-decomposed ERI, the input argument associated with the Cholesky instance (here \texttt{"abcd"}) is translated to the internal Cholesky representation, that is, \texttt{"xac,xbd"}.

Since \texttt{numpy.tensordot} only supports a summation over two multi-dimensional arrays at a time, we need to divide a tensor contraction containing more than two operands into appropriate intermediates.
Such a partitioning can be fully automatized, exploiting the \texttt{numpy.einsum\_path} function.
It proposes a contraction order of lowest possible cost for an \texttt{einsum} expression, taking into account the creation of intermediate arrays.
The resulting \texttt{tensordot\_helper} function has the following logic
\begin{lstlisting}[language=Python]
def tensordot_helper(subscripts, *operands):
    # split subscripts in input and output strings
    inscripts, outscript = split subscripts wrt "->"

    # sanity check
    if contraction cannot be performed:
        raise ArgumentError(error message)

    # get the optimial path
    path, _ = numpy.einsum_path(subscripts, *operands)

    # sometimes einsum_path does not return a list of two-element tuples
    if no optimized path has been found:
        path = provide some path

    # loop over all steps in the path
    for step in path[1:]:
        # take 1st and 2nd operands and corresponding subscripts
        op0, op1 = take the proper arrays from the operands list
        # Find summation axes
        axis0 = list containing axis indices to sum over for op0
        axis1 = list containing axis indices to sum over for op1
        axis = (axis0, axis1)
        # find default outscript
        outscript_ = store shape of intermediate
        # do contraction using numpy.tensordot
        outmat = numpy.tensordot(op0, op1, axes=axis)
        # update the list of operands and subscripts used in the next interation
        operands.append(outmat)
        scripts.append(outscript_)
    # do transposition if required.
    if outscript does not have the proper shape:
        trans = tuple how to transpose the output array
        outmat = outmat.transpose(trans)
    # return result of the tensor contraction
    return outmat
\end{lstlisting}
The \texttt{subscripts} argument is a string specifying the contraction using the \texttt{numpy.einsum} notation, while all multi-dimensional input arrays are stored in the \texttt{operands} argument.
We assume that the ERI corresponds to the leading subscripts.
If a tensor contraction cannot be performed, we transition to \texttt{opt\_einsum} or \texttt{numpy.einsum(..., optimize="optimal")} and an \texttt{ArgumentError} is raised.
In case of too large intermediates, the \texttt{tensordot\_helper} function is replaced by a function call containing selected hand-optimized tensor contraction operations.
The implemented flow of contraction operations (\texttt{CuPy}--\texttt{numpy.tensordot}--\texttt{opt\_einsum}/ \texttt{numpy.einsum}) allows for an optimal usage of computational resources, hardware, and multithreaded implementations.
The reasons for the proposed operational flow are scrutinized in section~\ref{sec:cpu-gpu-comparison}.
When the performance of the used libraries is improved in future releases, the order of the contraction flavors can be adjusted to maximize efficiency without significantly interfering with the underlying source code in each wave function module.

In the following, all benchmark calculations adhere to the contraction flow mentioned above if not stated otherwise.
That is, the majority of tensor contractions are performed using the \texttt{numpy.tensordot} routine, while the bottleneck contractions are outsourced to the GPU.

\section{\texttt{CuPy} and batch-wise computations}\label{sec:cupy}
While, of course, the most performance one could achieve by designing a method specifically for the hardware to be calculated, we focus on accelerating the mathematical bottleneck operations by exporting the contractions in question to the GPU.
\begin{center}
    \begin{figure}[h!]
       \centering
        \includegraphics[width=\columnwidth]{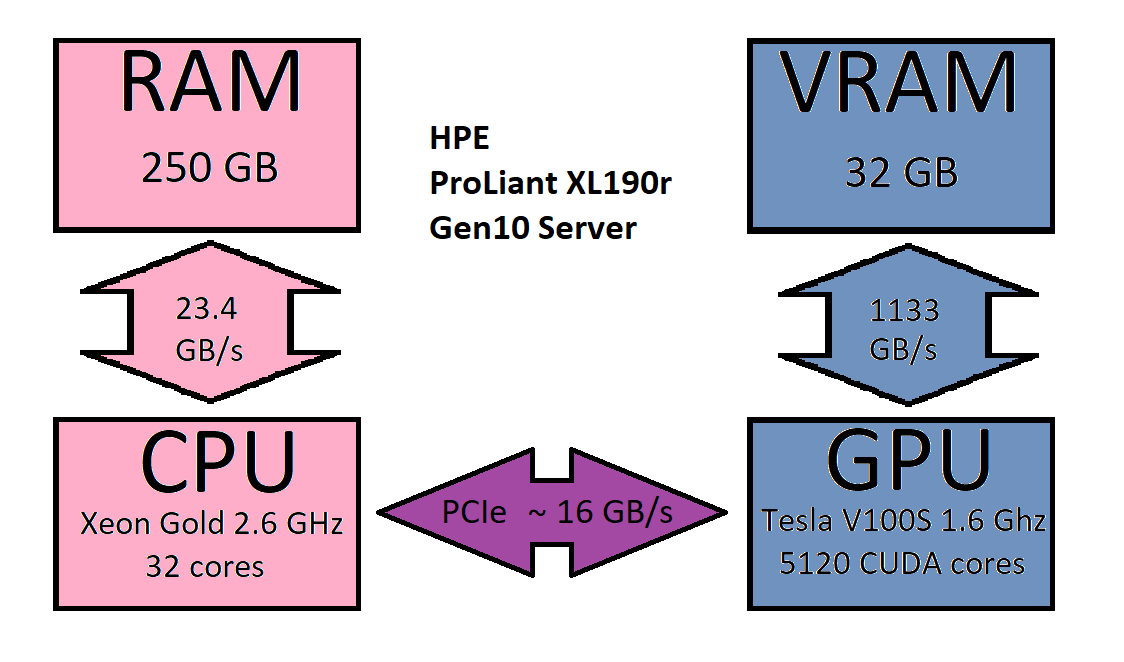}
        \caption{Schematic picture of RAM VRAM GPU and CPU and the data transfer rate between the components.}
        \label{fig:flowchart}
    \end{figure}
\end{center}
As proof of the concept, we exploit \texttt{CuPy}\cite{cupy_paper} for GPU-accelerated computing.
Specifically, it is written as a drop-in replacement for \texttt{NumPy}\cite{numpy-developers-blog}.
For medium- or large-sized molecules, we have to consider the size of the multi-dimensional arrays present in eq.~\eqref{eq:vfunction} and, therefore, the amount of data that needs to be processed and transferred to the Video RAM (VRAM). The principle of memory and processor communication is shown in Fig.~\ref{fig:flowchart}.
Due to their size, the underlying multi-dimensional arrays might be too big to be transferred, processed on the GPU, and transferred back.
Thus, a generic implementation, where a \texttt{NumPy} implementation is recycled as a \texttt{CuPy} implementation by replacing, for instance, \texttt{numpy.einsum} with \texttt{\texttt{cupy}.einsum} and taking into account host to device and device to host operations is impracticable or even impossible.

To calculate contractions on the GPU for realistic molecules and large basis sets, it is necessary to do the computations in batches.
In our batch-wise computing approach, the arrays on which the operations will be performed must be copied to the VRAM in chunks.
To maximize performance and utilize the massive number of CUDA cores, the size of the chunks has to be chosen as big as possible so that as few as possible cycles are needed.
To achieve reasonably-sized chunks of data to be processed in batches, we multiply the number of elements with their corresponding element size (in bytes) and sum up the required storage space for each multi-dimensional array involved.
If too large, the arrays are split, and the necessary memory is checked again.
This process is repeated till the split arrays fit into the video memory.
An example that shows the splitting, allocations, and de-allocations is shown in the following code example for the contraction \texttt{xac,xbd,ecfd->eafb}:
\begin{lstlisting}[language=Python]
t_new = numpy.zeros((t_old.shape[0], L_0.shape[1], t_old.shape[2], L_1.shape[1]))
start_e = 0
end_e = 0
# parts_d is the number of parts in which the dense array will be split
for e in range(0, parts_d):
    end_e += dense_e_chunk_lengths[e]
    start_a = 0
    end_a = 0
    # parts_c is the number of parts in which the Cholesky array will be split
    for a in range(0, parts_c):
        end_a += chol_chunk_lengths_0[a]
        start_b = 0
        end_b = 0
        # parts_c is the number of parts in which the Cholesky array will be split
        for b in range(0, parts_c):
            end_b += chol_chunk_lengths_1[b]
            
            # Batches of Cholesky arrays are copied to GPU Memory (VRAM)
            chol_0 = cupy.array(numpy.array_split(L_0, parts_c, axis=1)[a])
            chol_1 = cupy.array(numpy.array_split(L_1, parts_c, axis=1)[b])

            # calculating batch of xac,xbd->acbd on GPU
            result_temp = cupy.tensordot(chol_0, chol_1, axes=(0, 0))

            # deallocation
            del chol_0, chol_1
            cupy.get_default_memory_pool().free_all_blocks()
            
            # Batches of the vector function array is copied to VRAM
            operand = cupy.array(numpy.array_split(t_old, parts_d, axis=0)[e])

            # calculating batch of acbd,ecfd->abef on GPU
            result_temp_2 = cupy.tensordot(
                result_temp, operand,axes=([1, 3], [1, 3])
            )

            # deallocation
            del operand, result_temp
            cupy.get_default_memory_pool().free_all_blocks()

            # calculating batch of the transposition abef->eafb on GPU
            result_part = cupy.transpose(result_temp_2, axes=(2, 0, 3, 1))

            # deallocation
            del result_temp_2
            cupy.get_default_memory_pool().free_all_blocks()

            # result arrays are copied back to the memory (RAM)
            t_new[
                start_e:end_e, start_a:end_a, :, start_b:end_b
            ] = result_part.get()

            # deallocation
            del result_part
            cupy.get_default_memory_pool().free_all_blocks()

            start_b = end_b
        start_a = end_a
    start_e = end_e
\end{lstlisting}

\section{Numerical results}\label{sec:cpu-gpu-comparison}
\subsection{Software specifications and computational details}
All calculations were performed on the CentOS 7 operating system using, if not mentioned otherwise, the \texttt{CUDA compilation tools} v12.1.105, \texttt{intelpython} v3.9 (referred to as \texttt{Python} in the following), \texttt{Intel OneAPI} v2023.1, and \texttt{CuPy} v12.0.0.
Furthermore we employed \texttt{NumPy} v1.23.5, \texttt{opt\_einsum} v3.3.0, and \texttt{PyBEST} v1.4.0dev0.
The hardware on which the computations were performed is gathered in section S1 of the Supporting Information.
Note that for the benchmark data shown in Fig.~\ref{fig:einsum_vs_td-memory}, we used \texttt{Intel Python} 3.7, \texttt{Intel OneAPI} v2021.3, \texttt{PyBEST} v1.3.0, and \texttt{NumPy} v1.21.2.

The molecular structure of the L0 dye was optimized in the ORCA 5.0.3 software package~\cite{orca-2012, orca-2018, orca-2020, orca-2022} using the B3LYP~\cite{b3lyp, b3lyp_becke} exchange--correlation functional and the cc-pVTZ basis set.~\cite{basis_dunning}
The resulting structural parameters are provided in section S2 of the Supplementary Information.
That molecular structure is subsequently used in the orbital-optimized pair coupled cluster doubles (pCCD~\cite{limacher_2013, oo-ap1rog, tamar-pccd, pawel-pccp-geminal-review-2022, ap1rog-jctc, piotrus_mol-phys}) augmented with the linearized coupled cluster singles and doubles (pCCD-LCCSD) correction~\cite{ap1rog-lcc}, and the conventional coupled-cluster singles and doubles (CCSD) approach, as implemented in the \texttt{PyBEST} software package.~\cite{pybest-paper}
In all \texttt{PyBEST} calculations for the L0 dye, we utilized the Cholesky linear algebra factory with a threshold of $10^{-5}$ for the ERI.
In all benchmark calculations concerning timings and memory requirements, the Cholesky vectors are random arrays, where we assume a size of $5K^3$ with $K$ corresponding to the number of basis functions.
This roughly corresponds to a Cholesky cutoff threshold of $10^{-5}$ in actual molecular calculations.
\subsection{Comparison between CPU and GPU-accelerated implementations}
To be able to make assumptions about the benefit of offloading computations to the GPU, it is reasonable to study how effectively different functions described in section~\ref{sec:python-cc-vector} perform compared to each other.
In the following, the computation times of \texttt{numpy.tensordot}, \texttt{opt\_einsum}, and their CPU multicore processing behavior are investigated and compared with \texttt{CuPy}'s \texttt{tensordot}.
As mentioned above, a very handy way for implementing tensor contractions is \texttt{numpy.einsum} or \texttt{opt\_einsum}, which feature a similar syntax.
We should note that although \texttt{numpy.einsum} and \texttt{opt\_einsum} have similar performance if two operands are contracted with each other, this is not the case anymore if the list of operands contains several multi-dimensional arrays.
In the latter case, \texttt{opt\_einsum} may be superior to \texttt{numpy} in terms of computing time by several orders of magnitude, primarily due to the parallelization of the underlying lower-level \texttt{opt\_einsum} routines.
Thus, we only show benchmark results for \texttt{opt\_einsum} in this work.
Furthermore, we investigate three tensor contractions, namely \texttt{"abcd,ecfd->eafb"}, \texttt{"abcd,edfc->eafb"}, and \texttt{"abcd,ecfd->efab"}.
The first contraction (\texttt{"abcd,ecfd->eafb"}) corresponds to the bottleneck term of eq.~\eqref{eq:vfunction}, while the second one is the associated exchange term, which reads
\begin{equation}\label{eq:vfunction2}
    \sum_{xcd} L_{ad}^x L_{bc}^x t_{ij}^{cd}.
\end{equation}
Although the exchange part of the formal bottleneck contraction is not present when working in a spin-free representation, we benchmark this contraction for reasons of completeness, in case a spin-dependent implementation is sought.
The third contraction \texttt{"abcd,ecfd->efab"} recipe is used in two additional terms of the CCSD vector function.
These are one term involving an intermediate $\tilde{t}_{ijkl}$,
\begin{equation}\label{eq:vfunction3}
    \tilde{t}_{ijkl} = \sum_{xcd} L_{kc}^x L_{ld}^x t_{ij}^{cd},
\end{equation}
which is an operation of $\mathcal{O}(xo^4v^2)$ complexity, and another one comprising the ERI of $\mathcal{O}(ov^3)$,
\begin{equation}\label{eq:vfunction4}
     \tilde{t}_{ijbk} = \sum_{xcd} L_{bd}^x L_{kc}^x t_{ij}^{cd},
\end{equation}
with computational complexity of $\mathcal{O}(xo^3v^3)$.
Note that exporting eq.~\eqref{eq:vfunction3} to the GPU is merely a byproduct of eq.~\eqref{eq:vfunction4} as both contraction can be written using the same subscript.

\begin{figure}[!htb]
    \centering
        \centering
        \includegraphics[width=\columnwidth]{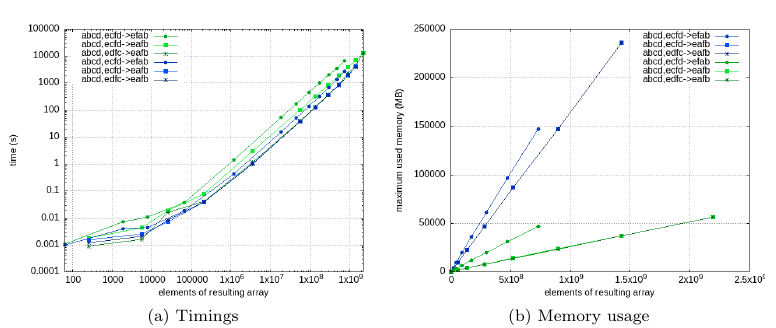}
        \caption{Timings (a) and memory usage (b) of \texttt{numpy.tensordot} and \texttt{opt\_einsum} with respect to the size (number of elements) of the largest input array. In both plots, the greenish colors show \texttt{tensordot} and blueish colors show \texttt{opt\_einsum} computations using 1 CPU core.}
        \label{fig:einsum_vs_td-memory}
\end{figure}

Compared to (our version of) \texttt{numpy.tensordot}, the internal optimization leads to a speed-up of factor 2 as indicated by Fig.~\ref{fig:einsum_vs_td-memory}(a).
The reason we chose \texttt{numpy.tensordot} as the workhorse of all tensor contractions instead of \texttt{opt\_einsum} (or \texttt{numpy.einsum}), is the severe disadvantage of \texttt{opt\_einsum} regarding memory efficiency.
This drawback becomes evident in the peak memory usage displayed in Fig.~\ref{fig:einsum_vs_td-memory}(b).
The internal optimization and the creation of intermediate arrays (for speed-up) lead to a factor 3 faster-growing memory consumption.
That disqualifies \texttt{opt\_einsum} as a generic option primarily because we do not want to and often cannot constrain the code to smaller problem sizes.
Therefore, we employ \texttt{numpy.tensordot} as the default contraction flavor if \texttt{Nvidia} \texttt{CUDA} is not available.
We should note that \texttt{opt\_einsum} features other arguments that limit the memory peak to a specific size.
However, this feature comes at the cost of computing time.
Specifically, a user-defined limit of the memory peak significantly deteriorates the speed of the numerical operations, which renders \texttt{opt\_einsum} impractical for large-scale tensor contractions.

\begin{figure}[!htb]
    \centering
        \includegraphics[width=\columnwidth]{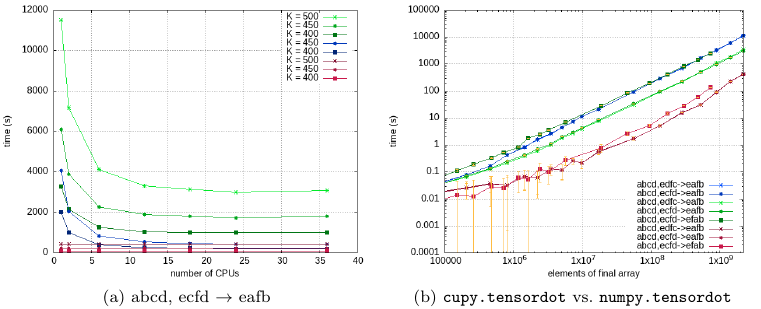}
        \caption{Computing times for CPU and GPU-accelerated implementations. (a) Timings for the bottleneck contraction \texttt{abcd,ecfd->eafb}. The greenish colors show \texttt{numpy.tensordot}, the blueish colors correspond to \texttt{opt\_einsum}, while reddish colors indicate \texttt{cupy.tensordot} results. $K$ is the total number of basis functions. (b) Comparison of GPU and CPU computing times for different basis set sizes $K$. The \texttt{numpy.tensordot} results for calculations with 1 CPU thread are shown in blueish colors and for 36 CPU threads in greenish colors. The reddish colors show the \texttt{cupy.tensordot} results. The error bars in orange are determined for an average of 10 runs and show the standard deviation. For the number of elements of the final array, see the description in the main text.}
        \label{fig:cupy_vs_td-timings}
\end{figure}

Fig.~\ref{fig:cupy_vs_td-timings}(a) summarizes timings for computations of the contraction \texttt{"abcd,ecfd->eafb"} (or \texttt{"xac,xbd,ecfd->eafb"} if the Cholesky vectors are explicitly mentioned) plotted over the number of CPU cores for various tensor contraction flavors and problem sizes $N$.
The greenish colors show \texttt{numpy.tensordot}, the blueish colors \texttt{opt\_einsum}, while reddish colors indicate \texttt{cupy.tensordot} timing results for different numbers of CPU cores.
Specifically for Fig.~\ref{fig:cupy_vs_td-timings}(a), the problem size is given by $N = o^2v^2$ because we investigate the following problem setup with dimensions $(xv^2,xv^2,o^2v^2\rightarrow o^2v^2)$, where $v$ and $o$ are the number of virtual and occupied orbitals respectively.
Their sizes have been set according to the relation $v=\mathrm{int}(\frac{3}{4}K)$ and $o=K-v$, where $K$ is the total number of basis functions.
We should mention that the data for $K = 500$ using \texttt{opt\_einsum} could not be obtained due to technical problems as the memory consumption/memory peak exceeds the available physical memory of the computing node.
Overall, the \texttt{opt\_einsum} calculations are a factor $3$ faster than the corresponding \texttt{numpy.tensordot} variants but limited by their memory peak.
Note, however, that the for-loop-based \texttt{numpy.tensordot} variant is roughly a factor of 2 slower compared to the contraction scheme where the $v^4$ intermediate is formed.
Thus, \texttt{opt\_einsum} and \texttt{numpy.tensordot} are comparable in performance, with the former being modestly faster.

Fig.~\ref{fig:cupy_vs_td-timings}(a) further highlights that the benefit of performing a computation on a larger number of cores shrinks very rapidly.
The timings reach a plateau at about 8 to 10 cores.
Judging by the numbers, the contraction functions do not really benefit from the usage of more than 10 cores, with 4 cores being the most reasonable amount, assuming the computing time is limited.
The bigger the basis set or problem size the more benefit one gets from utilizing a higher number of cores.
The {\texttt{cupy.tensordot} timings are also shown in Fig.~\ref{fig:cupy_vs_td-timings}(a) for a direct comparison.
We should note that they are independent of the number of CPU cores (as the mathematical operations are performed on the GPU) and faster than both the \texttt{NumPy} and \texttt{opt\_einsum} alternatives.

\begin{table*}[!tb]
  \begin{tabular}{lrrrl}
     & \multicolumn{3}{c}{$N$} & \\
    $K$ &300&400&500\\
    \hline
    \texttt{abcd,ecfd->eafb}&284 765 625&900 000 000&2 197 265 625&$o^2v^2$\\
    \texttt{abcd,edfc->eafb}&284 765 625&900 000 000&2 197 265 625&$o^2v^2$\\
    \texttt{abcd,ecfd->efab}&94 921 875 &300 000 000&732 421 875&$o^3v$
  \end{tabular}
  \caption{Problem sizes $N$ determined for different basis set sizes $K$ for the selected contractions benchmarked in this work. The last column indicates the size of the output array.}\label{tbl:timing}
\end{table*}
Fig.~\ref{fig:cupy_vs_td-timings}(b) compares the required computing time of CPU and GPU-accelerated contractions.
The results corresponding to GPU and CPU timings are taken as an average of over 10 runs.
Furthermore, the CPU data was obtained from computations exploiting 1 and 36 CPU cores, respectively.
Fig.~\ref{fig:cupy_vs_td-timings}(b) contains 3 different datasets, namely, one for each investigated contraction \texttt{"abcd,ecfd->eafb"}, \texttt{"abcd,edfc->eafb"}, and \texttt{"abcd,ecfd->efab"}.
Note that the notation is internally translated to the Cholesky vectors by replacing the \texttt{"abcd"} part of the string by \texttt{"xac,xbd"}.
Thus, we always have three input operands in each tensor contraction operation.
We should stress that the timings corresponding to all three contraction subscripts are very close to each other making them almost indistinguishable from each other on the plot.
For reasons of comparability, the x axis shows the problem size, which is given by the size of the resulting (output) tensor.
Table~\ref{tbl:timing} summarizes the different problem sizes with respect to the contraction subscript, namely, $N = o^2v^2$ for \texttt{"abcd,ecfd->eafb"} and \texttt{"abcd,edfc->eafb"} and $N = o^3v$ for the contraction \texttt{"abcd,ecfd->efab"}, respectively.
The standard deviation for the timings of larger problem sizes is about $1\%$ of the average value, while it increases to approximately $10\%$ for smaller problem sizes.
The elements of the arrays used to perform the benchmark calculations were randomly generated.
We attribute the larger standard deviation for smaller problem sizes to the higher impact a randomly generated array (e.g., very small numbers) can have, when among fewer elements in an operation.

All in all, we observe an order of magnitude reduction of computing time for the \texttt{CuPy} implementation computed on the \texttt{NVIDIA} Tesla V100S PCIe 32GB (rev 1a) compared to our \texttt{NumPy} implementation computed on 36 CPU cores for 3 different tensor contractions encountered in the CCSD working equations.

\subsection{A case study --- a sensitizer molecule}
To check the overall speed-up in actual chemical problems, we will perform example calculations with and without GPU acceleration.
Specifically, we will scrutinize the impact on the overall computing time when offloading just a given set of contractions to the GPU.
2-Cyano-3-(4-N,N-diphenylaminophenyl)-trans-acrylic acid, commonly referred to as L0 dye,\cite{l0_dye} has been chosen as the candidate for this performance test. 
This organic dye was designed to be a potential sensitizer in dye-sensitized solar cells, making it a viable alternative to ruthenium-type dyes.
The size of the L0 dye allows us to benchmark the Python-based hybrid CPU--GPU implementation for a chosen parameter set.

\begin{center}
    \begin{figure}[h!]
       \centering
        \includegraphics[width=\columnwidth]{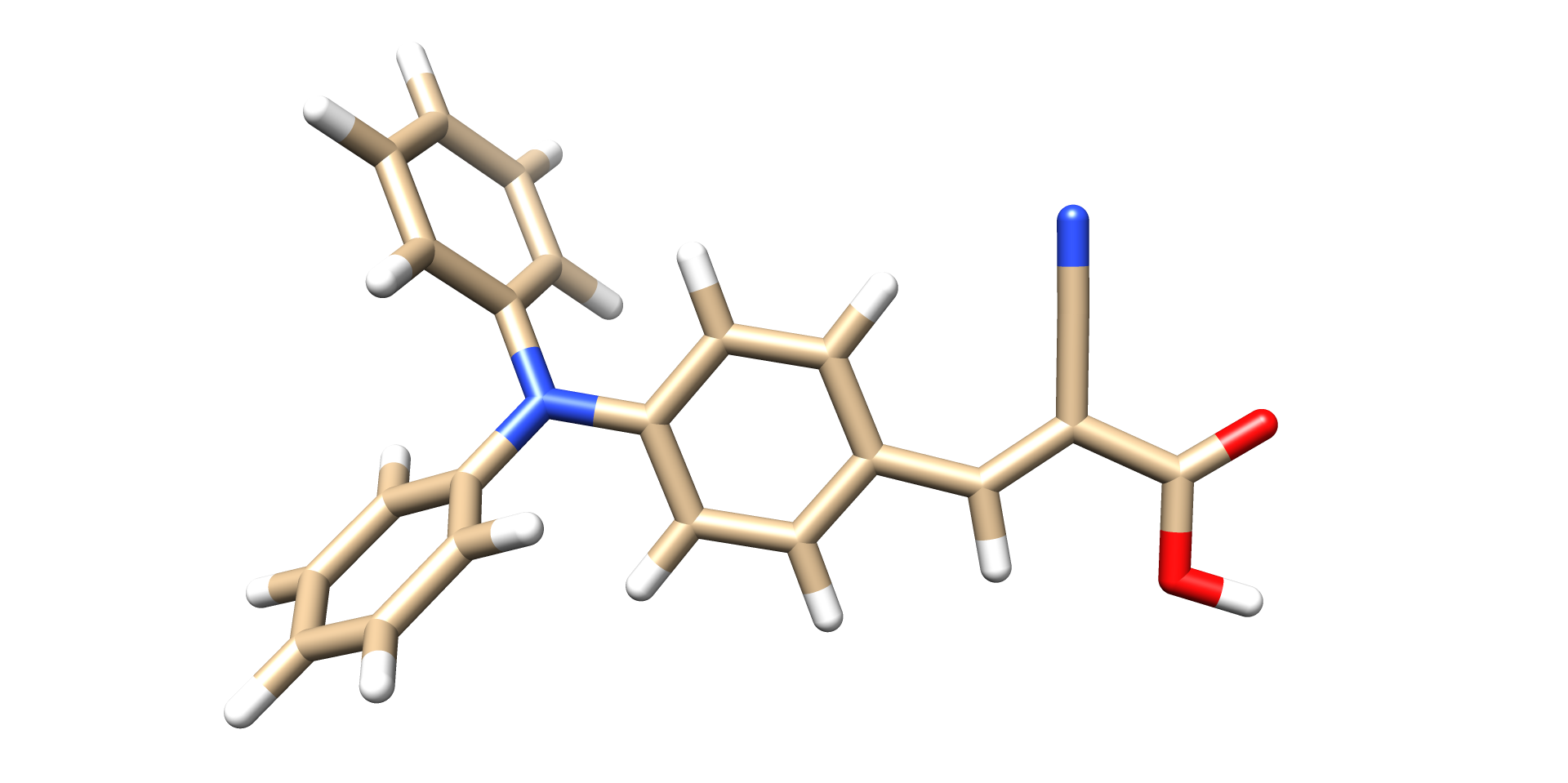}
        \caption{Molecular structure of the L0 dye relaxed at the B3LYP/cc-pVTZ level of theory.}
        \label{fig:structure}
    \end{figure}
\end{center}
\begin{table*}
  \begin{tabular}{llllll}\hline\hline
    && \multicolumn{2}{c}{CCSD}  & \multicolumn{2}{c}{pCCD-LCCSD}\\ \cline{3-6}
    &&\texttt{NumPy}&\texttt{CuPy}&\texttt{NumPy}&\texttt{CuPy}\\
    \hline
 Timings &      Average iteration step (CPU+GPU)&  2754.21  &  918.25   &   2692.91   & 813.80 \\  
         &      Average iteration step (CPU)	   &  2754.21  &  795.28   &   2692.91   & 680.00 \\
         &      Average iteration step (GPU)	   &    --     &  122.98   &     --      & 133.80 \\
         &      Vector function step (CPU+GPU)  &  2600.68  &  766.98   &   2541.80   & 662.38 \\
         &      Vector function step (CPU)      &  2600.68  &  644.00   &   2541.80   & 528.58 \\
         &      Vector function step (GPU)      &    --     &  122.98   &     --      & 133.80 \\
         &      Bottleneck contractions         &  1956.68  &  122.98   &   2013.23   & 133.80 \\ \hline
Speed-up &      Bottleneck contractions         &\multicolumn{2}{c}{16} &\multicolumn{2}{c}{15}\\
         &      Vector function step            &\multicolumn{2}{c}{3.4}&\multicolumn{2}{c}{3.8}\\\hline\hline
    \caption{Timings [s] and speed-up factors for the CPU and hybrid CPU--GPU implementation for selected CC calculations using a cc-pVDZ basis set (444 basis functions). Iteration step time indicates the time [s] required for one CC iteration step. It contains the evaluation of the vector function, the update of the CC amplitudes, and the evaluation of the CC energy expression. All timings correspond to differences in epoch times. Average iteration step: mean value for the time of one CC iteration averaged over 4 steps. Vector function step: time of the CPU/GPU part of the vector function averaged over 4 steps. Bottleneck contractions: time for all bottleneck contractions investigated in this work, that is, those exported to the GPU, averaged over 4 steps.}\label{tbl:dye}
  \end{tabular}
\end{table*}
We tested our CCSD and pCCD-LCCSD implementations by exploiting a cc-pVDZ basis set in our benchmark calculations.
All calculations were performed with 36 parallel threads.
Table~\ref{tbl:dye} compares different CPU timings in seconds for \texttt{cupy.tensordot} to the timings of our \texttt{numpy.tensordot} variant.
As highlighted in Table~\ref{tbl:dye}, one iteration step of the vector function takes around 2600 seconds on the CPU, while the corresponding function requires only 770 seconds to be evaluated in the case of the hybrid CPU--GPU variant.
Note that the average iteration step time shown in Table~\ref{tbl:dye} includes the evaluation of the vector function, the update of the CC amplitudes, and the evaluation of the CC energy expression, which is similar for the CPU and hybrid CPU--GPU implementation as those operations are not offloaded to the GPU.
The difference in the computing times between the CPU and CPU--GPU implementation of the vector function is the time used by the contractions that were offloaded to the GPU, namely around 1950 s for the bottleneck contractions per CC iteration step compared for the pure CPU variant.
In contrast, the bottleneck contractions offloaded to the GPU require only about 123 s per CC (vector function) iteration step, which is a speed-up of approximately a factor of 16.
Comparing the resulting iteration times for the vector function evaluation of the CPU-based \texttt{numpy.tensordot} implementation to the CPU--GPU hybrid variant, $2600.68/766.98\approx3.4$, we obtain a total speed-up of approximately a factor of 3.
Furthermore, while the bottleneck contractions require about 74\% of the computing time per iteration step on the CPU, it drops down to approximately 16\% of the computing time per iteration using a CPU--GPU hybrid approach.

We observe similar speed-ups for the pCCD-LCCSD method.
We should note that our generic GPU implementation also offloads the contraction \texttt{"abcd,cedf->aebf"} to the GPU in addition to the bottleneck operation \texttt{"abcd,ecfd->eafb"}.
The third bottleneck operation \texttt{"abcd,ecfd->efab"} corresponds to disconnected $\hat T_2$ terms and does not show up in the pCCD-LCCSD vector function.
This additional tensor contraction corresponds to a term of $\mathcal{O}(o^4v^2)$.
As expected, the evaluation time of the pCCD-LCCSD vector function takes less time compared to the CCSD implementation as we exclude the majority of the disconnected terms.
Offloading to the GPU reduces the computing time for the selected bottleneck operations from about 2000 s to 133 s, which corresponds to a speed-up factor of approximately 15.
The overall speed-up for one evaluation of the pCCD-LCCSD vector function drops to a factor of $2541.80/662.38\approx3.8$.
On average, the bottleneck contractions take about 20\% of the computing time of the vector function per iteration step for the CPU--GPU hybrid implementation, while the corresponding time increases to 80\% for the CPU variant, which is similar to our CCSD example.

We conclude that offloading the slowest contraction of the form of eq.~\eqref{eq:vfunction} to the GPU already leads to a significant acceleration, shifting the bottleneck to another set of terms of the CC working equations.
Additional speed-up can be obtained by offloading other speed-determining tensor contractions to the GPU.

\section{Conclusions and outlook}\label{sec:conclusions}
Current trends in scientific programming heavily rely on Python-based implementations~\cite{python-in-chemistry}, which are easy to code but need to be more easily scalable to high-performance computing architectures. 
At the same time, the potential end-users of these codes would like to work on supercomputers to solve problems as large as possible without a profound knowledge of high-performance optimizations.
To meet those needs for quantum chemistry problems, we analyzed the limitations of quantum chemistry methods written in Python, where some trade-off between the memory and CPU time has to be made. 
We showed how to use the existing Python libraries to speed up quantum chemical calculations and provided numerical evidence, including comparisons between CPU and GPU. 

A common and reasonable practice to speed up an implementation is to identify the so-called bottleneck operations and focus on optimizing these routines.
In this work, we focused on the bottlenecks of CCSD-type methods, which are a given set of tensor contractions, and their translation into Python code using various third-party libraries.
Specifically, we scrutinized computing timings and memory consumption, comparing \texttt{numpy.tensordot} and \texttt{opt\_einsum} calculations.
We found that \texttt{opt\_einsum} computes about a factor 2.5 faster than a modified version of \texttt{numpy.tensordot} on a CPU but is limited due to tremendous memory peaks and, therefore, not a universal candidate when large system sizes are considered.
Furthermore, we have rewritten selected \texttt{numpy.tensordot} routines imposing one \texttt{for} loop to prevent the construction of large intermediate arrays.
In general, this \texttt{for}-loop implementation is responsible for the drop in efficiency (speed) of \texttt{numpy.tensordot} compared to \texttt{opt\_einsum} by approximately a factor of 2.
Nonetheless, the gain in memory reduction outweighs the decrease in speed.
Although third-party libraries are convenient to interface and easy to use, special attention has to be paid to possible intermediates created during the evaluation of, for instance, tensor contractions.
Both \texttt{opt\_einsum} and various \texttt{NumPy} linear algebra routines may create several intermediate arrays to increase speed-up.
These memory outbursts may impede a black-box implementation that is computationally feasible for large-scale problems.

A promising alternative for Pythonic large-scale computing is GPUs. We implemented this set of bottleneck tensor contractions to be calculated on the GPU using \texttt{CuPy}, an application programming interface (API) to \texttt{Nvidia CUDA} for \texttt{Python}.
Due to the size of the arrays, it is necessary to perform the computations on the GPU batch-wise by splitting the overall tensor contractions into smaller-sized problems that fit into the video memory (VRAM). Utilizing this implementation and performing benchmark calculations of the bottleneck contractions showed that a single GPU already leads to a factor of 10 speed-up compared to our \texttt{NumPy}-based methods using 36 CPU cores.
This factor 10 speed-up of the contraction routines only translates to an overall speed-up of factor 3 as the bottleneck of the CC vector function evaluation has shifted to a collection of terms of similar scaling, which are still evaluated on the CPU.
Most importantly, our timing benchmark results are encouraging and again prove the potential of GPU utilization in Python-based computational chemistry.

Subsequently, we will identify the ``new'' bottleneck operations in tensor contractions and export them to the GPU for additional potential speed-up.
Better utilization of the tensor cores could also lead to improvement, as good speed-ups are reported in developer forums, where FP32 input/output data in DL frameworks and HPC can be accelerated, running ten times faster than V100 FP32 FMA operations.~\cite{nvidia-developers-blog}
Yet, additional speed-up for quantum chemistry problems is expected on the NVIDIA A100 hardware~\cite{rank-redued-ccsd-gpu-martinez-jcp-2022}, which has more GPU cores and up to 80 GB of memory. 
Generally, direct back-end CUDA implementation in \texttt{C++} also offers plenty of room for improvements like optimization of the array structure or concurrent computing.
Alternative routes to export Python code to the GPU are, for instance, offered by the Intel oneAPI toolkits.
Finally, multi-GPU utilization~\cite{multinode-gpu-eri-matinez-jctc-2022} could further improve the already tremendous speed-up of factor 10.

\begin{acknowledgement}
The research leading to these results has received funding from the Norway Grants 2014--2021 via the National Centre for Research and Development.
P.T.~acknowledges financial support from the OPUS research grant from the National Science Centre, Poland (Grant No. 2019/33/B/ST4/02114). P.T.~acknowledges the scholarship for outstanding young scientists from the Ministry of Science and Higher Education. M.G.~acknowledges financial support from a Ulam NAWA -- Seal of Excellence research grant (no.~BPN/SEL/2021/1/00005).
\end{acknowledgement}


\scriptsize
\bibliography{rsc}

\end{document}